\documentclass[pra,twocolumn,epsfig,rotate,superscriptaddress,showpacs]{revtex4}
\usepackage[dvips]{graphics}
\usepackage{graphicx}
\usepackage{epsfig}
\usepackage{overpic}
\usepackage{amsfonts}
\usepackage{amssymb}
\usepackage{amsmath}
\newcommand{\cK}{\ensuremath{\mathcal{K}}}
\newcommand{\cL}{\ensuremath{\mathcal{L}}}
\newcommand{\cM}{\ensuremath{\mathcal{M}}}
\newcommand{\img}{\mathrm{i}}
\newcommand{\nte}{\mathrm{e}}
\newcommand{\nn}{\nonumber \\}

\usepackage{prettyref}
\usepackage{float}
\usepackage{amsmath}
\usepackage{subfigure}

\usepackage{amssymb}
\usepackage{esint}

\usepackage{amsmath,bm}
\usepackage{color}

\begin{document}
\title
{Boosting work characteristics and overall heat engine performance via shortcuts to adiabaticity: quantum and classical systems}

\author{Jiawen Deng}
\affiliation{Department of Physics, National University of Singapore, 117542,
Singapore}
\author{Qing-hai Wang}
\affiliation{Department of Physics, National University of Singapore, 117542,
Singapore}
\author{Zhihao Liu}
\affiliation{Department of Physics, National University of Singapore, 117542,
Singapore}
\author{Peter H\"{a}nggi}
\affiliation{Department of Physics and Centre for Computational
Science and Engineering, National University of Singapore, 117542,
Singapore}
\affiliation{Theoretische Physik I, Institut f\"{u}r
Physik, Universit\"{a}t Augsburg, D - 86135 Augsburg, Germany}
\author{Jiangbin Gong} \email{phygj@nus.edu.sg}
\affiliation{Department of Physics and Centre for Computational
Science and Engineering, National University of Singapore, 117542,
Singapore}

\date{\today}
\begin{abstract}

{Under a general framework, shortcuts to adiabatic processes are shown to be possible in classical systems}.
We then study the distribution function of the work done on a small system initially prepared at thermal equilibrium.  It is found that the work fluctuations can be significantly reduced via shortcuts to adiabatic processes. For example, in the classical case
 probabilities of having very large or almost zero
 work values are suppressed.
 In the quantum case negative work may be totally removed from the otherwise
 non-positive-definite work values.
 We also apply our findings to a micro Otto-cycle-based heat engine. It is shown that
 the use of shortcuts, which directly {enhances} the engine output power, can
 also increase the heat engine efficiency substantially, in both quantum and classical regimes.

\end{abstract}

\maketitle

\section{Introduction}
  Shortcuts to adiabatic processes (STA) constitute a timely topic of broad interest \cite{Rice,Berry,Nakamura,Chen1,Chen2,onofrio,NJP,Nature, PRLsuter,campoPRL,campo,chris,campo3}, with several experimental realizations reported recently \cite{NJP,Nature, PRLsuter}. One important question follows: are such STA unique in quantum mechanics?

In this work, we first develop a simple and {\it general} framework for {\it classical} STA.
 Analogous to the classical adiabatic theorem, the involved Hamiltonian is assumed to be integrable \cite{inte-note}. A generic control field to achieve
 classical STA
 is then found.   This is important because (i) classical STA may help to design quantum STA {(e.g., by quantizing the classical control field)} and (ii) STA may be more general and robust than previously thought. Indeed,
 applying our formalism to a parametric oscillator, the control Hamiltonian is precisely the classical limit of an early quantum result \cite{CampoJPB}.

 To make a new connection between STA and nonequilibrium statistical mechanics, we ask how STA impact the distribution function of the work done on a single system initially prepared in thermal equilibrium. We find that the work fluctuations can be significantly reduced. For example, both the long-tail part and the almost zero part of the work function
in a classical example are substantially suppressed, leading to a faster convergence towards Jarzynski's equality. Remarkably, in the corresponding
quantum case negative work values of the otherwise non-positive-definite quantum work may be  removed {\it completely}.  
 We then show how STA {implemented} in a prototype micro heat engine \cite{kosloff,Lutz} can increase engine efficiency and the power output at the same time.

  \section{Shortcuts to adiabatic processes: from quantum to classical systems}
    Consider first STA for a quantum system (also called transitionless driving in
    Ref.~\cite{Berry}) for a quantum system with a non-degenerate Hamiltonian $\hat H_0[\lambda(t)]$, parameterized by $\lambda(t)$, with the $n$-th instantaneous energy eigenstate given by $|n[\lambda(t)] \rangle$ possessing eigenenergy $E_n[\lambda(t)]$. {The} quantum adiabatic theorem states that, if $\lambda$ changes slowly enough, the system initially prepared on an eigenstate $|n[\lambda(0)]\rangle$ will continue to stay on the instantaneous energy eigenstate $|n[\lambda(t)] \rangle$, {\it i.e.},
\begin{equation}
\hat U(t,0)|n[\lambda(0)]\rangle \approx \nte^{\img\phi(t)} |n[\lambda(t)]\rangle,
\end{equation}
where $\hat U(t,0)$ is the unitary time evolution operator, and $\phi(t)$ includes a dynamical phase and a geometrical phase \cite{BerryPhase}.  This solution clearly indicates the invariance of population on each instantaneous eigenstate.
Such type of
time evolution can be accelerated by adding a control Hamiltonian $\hat{H}_C(t)$. In particular,
by reverse engineering \cite{Rice,Berry} one finds
\begin{equation}
\hat H_C(t) =  \img\hbar \frac{\partial U(t,0)}{\partial t} U^{\dagger}(t,0)- \hat H_0[\lambda(t)].
\end{equation}
This $\hat H_C$ assists an adiabatic process associated with $H_0$ (the process is certainly non-adiabatic with respect to the
full Hamiltonian $\hat H_0 +\hat H_C$), regardless of how fast $\lambda(t)$ varies.
Note that $\hat H_C \propto \dot{\lambda}$ \cite{Rice,Berry}.
In the limit $\dot{\lambda}\rightarrow 0$, $\hat H_C$ also approaches zero and the quantum adiabatic theorem is recovered.

We now show that classical adiabatic processes also have shortcuts \cite{notegong,chris}, and the physics is analogous to the above quantum picture. Though our treatment
applies to arbitrary multi-dimensional
integrable classical systems, for convenience
we consider here a time-dependent classical Hamiltonian $H_0[p,q,\lambda(t)]$ with one degree of freedom [$(p,q)$ are phase space variables].   To study classical adiabatic processes we further assume {that} $H_0[p,q,\lambda]$ at a fixed $\lambda$ can be written
as $\tilde H_0(I,\lambda)$: a function of the action variable $I$ but not a function of
the angle variable $\theta$ \cite{Goldstein}.  For a general time-dependent $\lambda(t)$,
 the correct Hamiltonian $K_0[I,\theta,t]$ in the action-angle representation is obtained by use of a time-dependent type-II {\it generating function} $F_2[I,q,\lambda(t)]$ between $(p,q)$ and $(I,\theta)$ \cite{Goldstein}, {\it i.e.},
\begin{eqnarray}
K_0[I,\theta,t] = \tilde H_0[I,\lambda] + \left[ \left({\partial F_2(I,q,\lambda)\over \partial \lambda}\right)_{I, q}\dot \lambda\right]\Biggr|_{q=q(I,\theta,\lambda)}.
\label{classicaladiabatic}
\end{eqnarray}
Classical adiabatic theorem then states that $I$ is approximately a constant if $\lambda$ is changing slowly.

To realize classical STA, {\it i.e.}, to have a constant $I$ with fast changes in $\lambda$, we construct a control Hamiltonian $K_C[I,\theta,t]$ on top of $K_0[I,\theta,t]$. One obvious but exact solution is
\begin{eqnarray}
K_C[I,\theta,t] = -\left[ \left({\partial F_2(I,q,\lambda)\over \partial \lambda}\right)_{I, q}\dot \lambda\right]\Biggr|_{q=q(I,\theta,\lambda)}.
\label{classicalcontrol}
\end{eqnarray}
The total Hamiltonian then becomes
\begin{eqnarray}
K = K_0 +K_C= \tilde H_0[I,\lambda],
\end{eqnarray}
which is $\theta$-independent, and as such $I$ is
a constant of motion {\it exactly}.  Note that $K_C \propto \dot{\lambda}$.
In the limit ($\dot\lambda\rightarrow 0$), {the} classical adiabatic theorem is recovered.
The theoretical result here is consistent with the above quantum STA result also because
$I$ is an analog of the quantum number $n$.


  \section{ Classical versus quantum work fluctuations}
  For a small system under a control protocol, the work done on a system fluctuates.    To study the impact of STA [which establishes
a constant $I$ (constant $n$) in classical (quantum) cases] on the work statistics,
we introduce work functions below and
advocate a useful scheme to connect STA with nonequilibrium statistical mechanics.

For a classical system $H_0[p,q,\lambda(t)]$ initially prepared at $(p_0,q_0)$, the inclusive work \cite{Jar,JarCR,rmp} during
a period of $0\leq t \leq \tau$
is given by
\begin{equation}
W_\tau = H_0[p(p_0,q_0,\tau),q(p_0,q_0,\tau),\lambda(\tau)]-H_0[p_0,q_0,\lambda(0)],
\end{equation}
where $[p(p_0,q_0,t), q(p_0,q_0,t)]$ represents a classical trajectory emanating from $(p_0,q_0)$.
The classical work  function $P^c(W)$ is defined as
\begin{equation}
P^c(W) =\int\rho(p_0,q_0)\delta[W-W_\tau(p_0,q_0)]dp_0dq_0,
\end{equation}
where $\rho(p_0,q_0)$ describes the initial statistical ensemble, {\it e.g.},
 the Gibbs distribution corrsponding to $H_0[\lambda(0)]$.  To have a fair comparison with the bare cases without  $H_C$, we propose a scheme
in which $\dot\lambda (0) =\dot\lambda(\tau)=0$ and hence
 \begin{equation}
 H_C(p,q,0)=H_C(p,q,\tau)=0.
  \end{equation}
That is, when evaluating the work, the control Hamiltonian $H_C$ does not directly affect the calculation because $H_C$ vanishes
in the beginning and in the end. On the other hand, the trajectories are affected by
$H_C$ and the work done to the system is now achieved by both a device achieving $\lambda(t)$ and by a field implementing $H_C$.
Interestingly, under our control scheme,
the
Jarzynski's equality \cite{Jar,rmp}, {\it i.e.},
\begin{equation}
\langle e^{-\beta W}\rangle= e^{-\beta\Delta F}
\end{equation}
 still holds for the same $\Delta F$
even with $H_C$. Note that here $\langle\cdot\rangle$ represents the thermal average, and $\Delta F$ is the free energy difference between a thermal equilibrium state of the final-state configuration (at the same temperature) and the initial thermal state.

A comparison between classical and quantum work fluctuations will be stimulating. To that end we adopt the two-time measurement definition of quantum work \cite{HanggiWork}, which is known as
$W = E_m^\tau-E_n^0 $
with $E_m$ and $E_n$ being the energy values upon energy measurement. The quantum work function $P^q(W)$ is then given by
\begin{equation}
P^q(W)=\sum_{m,n} P_n^0P_{n \rightarrow m}^\tau\delta[W-(E_m^\tau -E_n^0)],
\label{quantumP(W)}
\end{equation}
where $ P_n^0$ is the initial probability on $|n[\lambda(0)]\rangle$ and $P_{n \rightarrow m}^\tau$ is the transition probability between $|n[\lambda(0)]\rangle$ and $|m[\lambda(\tau)]\rangle$.
Adopting the same scheme as in classical cases {\it i.e.,} $\dot\lambda (0) =\dot\lambda(\tau)=0$, we have
\begin{equation}
\hat H_C(0)=\hat H_C(\tau)=0.
\end{equation}

 \section{Applications to a parametric oscillator}
  A parametric oscillator is feasible for experimental investigations \cite{kosloff,Lutz}.  We are thus motivated to consider a parametric oscillator with a time-dependent angular frequency $\omega(t)$. In the quantum version, the Hamiltonian is
  \begin{equation}
  \hat H_0(t) = {\hat p^2\over 2m} + {m\over 2} \omega^2(t)\hat q^2.
    \end{equation}
    The required $\hat H_C(t)$ {used} to realize quantum transitionless driving is found to be \cite{CampoJPB}
    \begin{equation}
    \hat H_C(t) = -{ \dot\omega \over 4\omega}(\hat q\hat p + \hat p\hat q),
\end{equation}
which is proportional to $\dot \omega$. Such a control Hamiltonian may be realized by considering
a magnetic field, whose vector potential $\hat{A}$ is proportional to $\hat q$ (this also effectively changes $\omega$). In the classical domain, the classical Hamiltonian is
\begin{equation}
H_0[p,q,\omega(t)] ={p^2\over 2m}+{m\over 2}\omega^2(t)q^2.
 \end{equation}
 Using our general result above
one finds
\begin{equation}
K_C[I,\theta,t]  = -{\dot \omega\over \omega}I\sin\theta \cos\theta.
\end{equation}
Detailed calculations in the Appendix \cite{notegong} show
\begin{equation}
H_C[p,q,t]\equiv K_C[I(p,q,t),\theta(p,q,t),t]  =  - {\dot \omega\over 2\omega} p q,
\end{equation}
which is precisely the classical limit of the quantum $\hat{H}_C$.

For work fluctuations we consider the classical case first.
 Because $\tilde H_0[I,\omega(t)]=\omega(t) I$ and we have set $\dot\omega(0)=\dot\omega(\tau)=0$, we find
$K[I,0]=\omega(0) I  $ and
$K[I,\tau]=\omega(\tau) I$, with $I$ being an exact constant of motion under $H_C$.
 The work expression then reduces to
\begin{eqnarray}
W_\tau &=& H_0[p(p_0,q_0,\tau),q(p_0,q_0,\tau),\omega(\tau)]-H_0[p_0,q_0,\omega(0)] \nonumber  \\ & = & K[I,\omega(\tau)]-K[I,\omega(0)] \nonumber
\\ &= & \Delta\omega I,
 \end{eqnarray}with $\Delta\omega \equiv \omega_f-\omega_i$ taken to be positive.
For simplicity we also denote $\omega_f\equiv\omega(\tau)$ and $\omega_i\equiv\omega(0)$.
Assuming that the system is initially prepared in a canonical distribution with inverse temperature $\beta$, i.e.,
\begin{equation}
\rho(I,\theta) = {\beta \omega_i \over 2\pi} e^{-\beta \omega_i I},
\end{equation}
we obtain
\begin{eqnarray}
P^c_{\rm ad}(W)= {\omega_i\beta\over\Delta\omega}\exp\left(-{\omega_i\beta W\over\Delta\omega}\right)\Theta(W),
\end{eqnarray}
with $\Theta$ being the unit step function.
Because this result is independent of details of $\dot\omega$, it does apply to conventional adiabatic cases as well. That is, the work function for STA remains identical with that in conventional adiabatic processes {\cite{normaladiabatic}}.

It is necessary and interesting to compare $P^c_{\rm ad}(W)$ with the work function $P^c_{\rm nonad}(W)$ of a process with the same time-dependence of $\omega(t)$ but without $H_C$. For the bare protocol we obtain
\begin{eqnarray}
P^c_{\rm nonad}(W)&=& {1\over\sqrt{\mu_+\mu_-}}\exp{\left[-{\mu_++\mu_-\over 2\mu_+\mu_-}W\right]} \nonumber \\
&\times & I_0\left[{\mu_+-\mu_-\over 2\mu_+\mu_-}W\right] \Theta(W),
\end{eqnarray}
where $I_\alpha(x)$ is the modified Bessel function of the first kind with parameter $\alpha$. Here
we have further assumed an (positive-valued) angular frequency $\omega(t)$ increasing monotonically during the protocol, with a simple proof
of the positive-definiteness of $W$ detailed in the Appendix.
The constant factors $\mu_\pm$ should be determined by specific realizations of $\omega(t)$. In particular,
in the sudden change limit (SL) of $\tau\rightarrow 0$,
$P^c_{\rm nonad}(W)$ reduces to
\begin{equation}
P^c_{\rm SL}(W)=\sqrt{1\over\pi W} \sqrt{\frac{\beta \omega_i^2}{\omega_{f}^2-\omega_i^2}}
\exp\left[-{\beta\omega_i^2\over \omega_{f}^2-\omega_i^2}W\right] \Theta(W).
\end{equation}
 Interestingly, $P^c_{\rm SL}(W)$ diverges at $W=0$ whereas $P^c_{\rm ad}(W)$ is always finite. Equally interesting, since we assume $\omega_f > \omega_i$ (without loss of generality), we have
 \begin{equation}
\frac{\omega_i^2}{\omega_{f}^2-\omega_i^2}<\frac{\omega_i}{2\Delta\omega}.
\end{equation}
So the exponential decay rate of $P^c_{\rm SL}(W)$, which is ${\beta\omega_i^2\over \omega_{f}^2-\omega_i^2}$, is less than half of the exponential decay rate of $P^c_{\rm ad}(W)$ for STA.
Two main impacts of STA on the work function are hence clear:
the probabilities for both very small and large work values are strongly suppressed.



\begin{figure}[t]
\begin{center}
\resizebox *{6.8cm}{5.1cm}{
\begin{overpic}[scale=1]{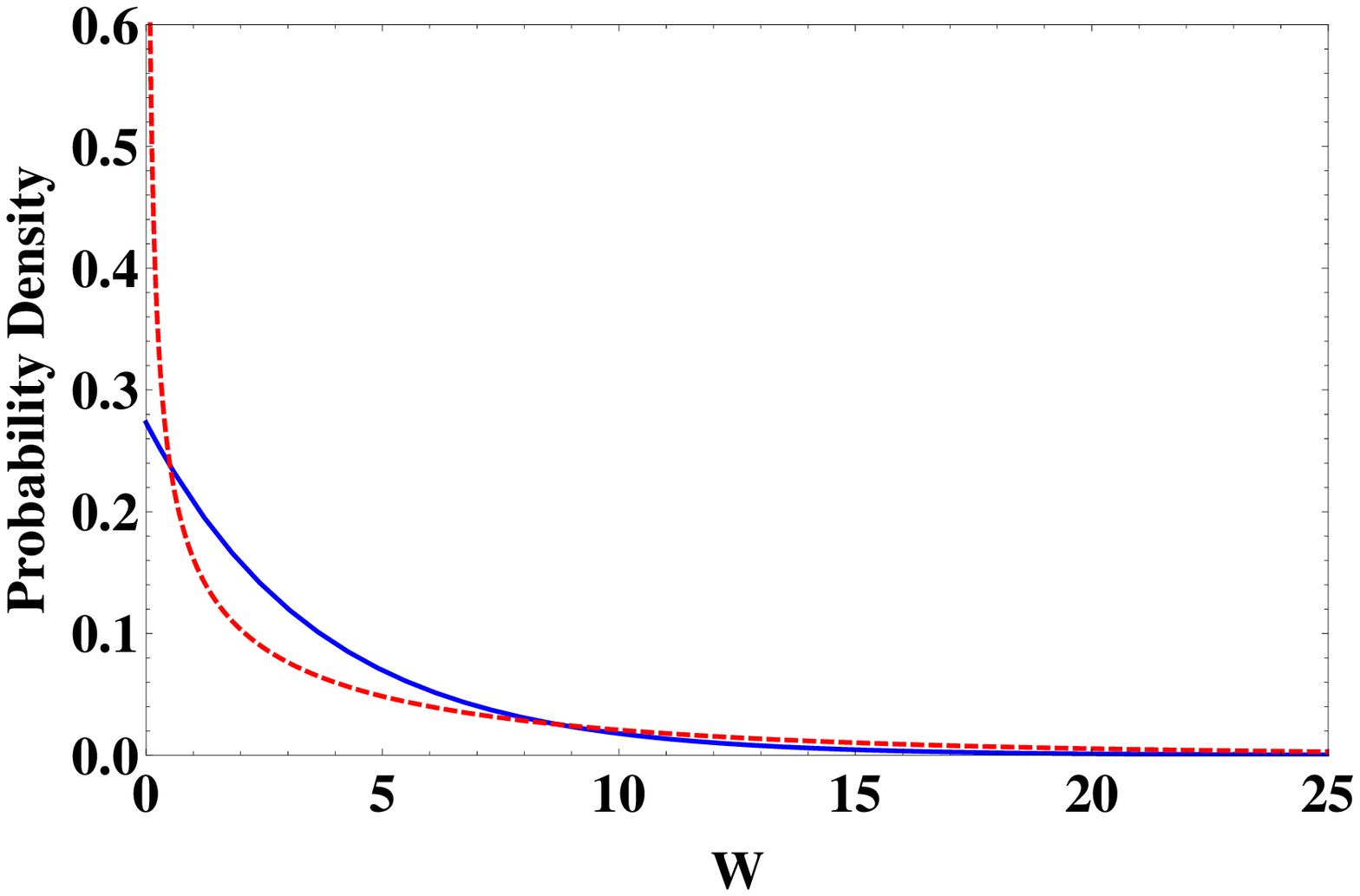}
  \put(38,25){%
    \includegraphics[scale=.8]%
    {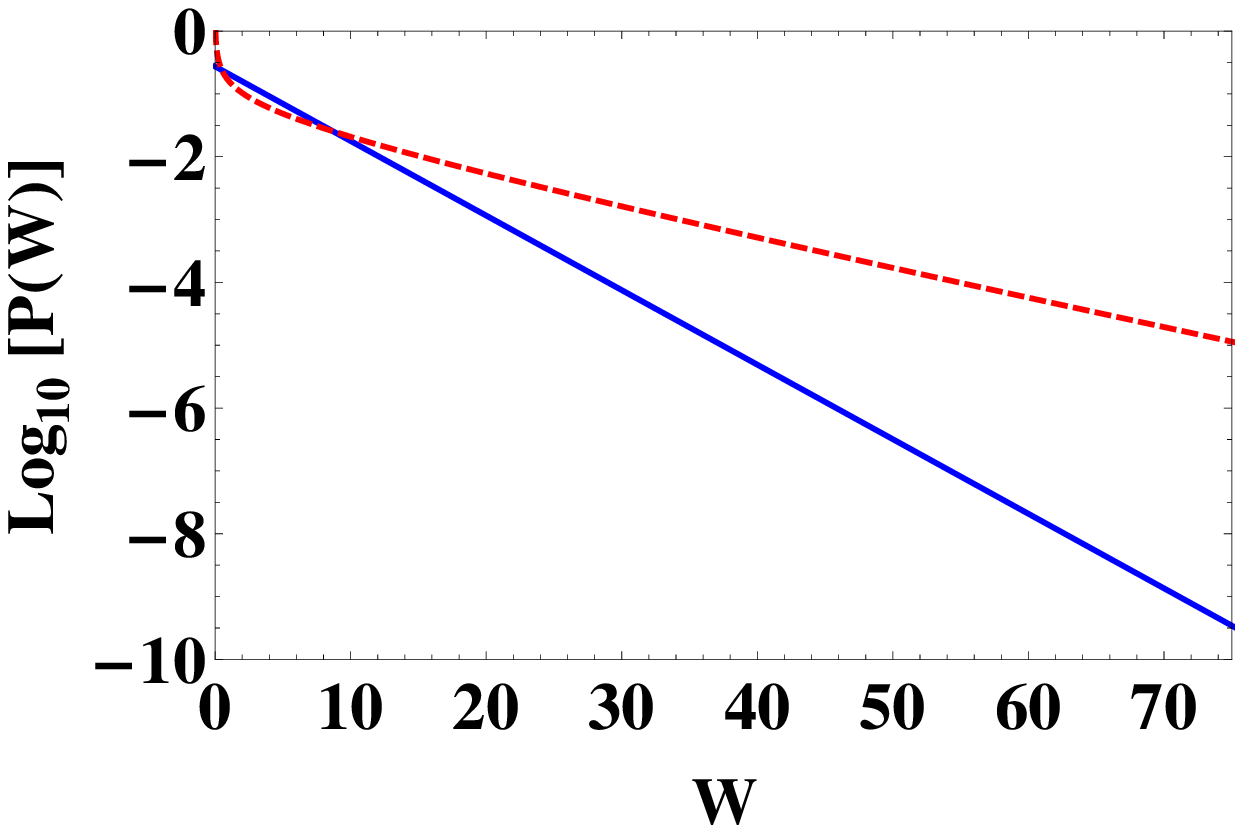}}
\end{overpic}}
\end{center}
\caption{(color online) Work function for a parametric oscillator with $\beta = 0.2$, $\omega_i=10$ and $\omega_f=10\sqrt{3}$, with all variables in scaled and hence dimensionless units. Solid (blue) line is for STA, with $\omega_i \tau=0.001$ and
$\omega(t)$ chosen to be $\omega(t)=\omega_i \sqrt{{a^2+1\over 2}-{a^2-1\over2}\cos (\pi{t\over\tau})}$, with $a=\omega_f/\omega_i=\sqrt{3}$.
Dashed (red) line is for a bare process of the same duration.  Same results are shown in the inset using a semi-log plot.}
\label{fig:subfig:P(W)}
\end{figure}



To illustrate and corroborate our analysis we present in Fig.~\ref{fig:subfig:P(W)} numerical results of the work function. The results agree with our theoretical calculations.
Indeed, compared with a bare protocol of the same duration, the STA case (i) suppresses the long tail of the work distribution and {(ii)} also significantly decreases the weights of almost zero work.  Quantitatively, in terms of mean work $\langle W\rangle$ and the standard deviation of work $\sigma(W)$,
\begin{equation}
\langle W\rangle_{\text{STA}}=\sigma(W)_{\text{STA}}=\frac{\Delta\omega}{\omega_i\beta}
 \end{equation}
 for STA, whereas for a nonadiabatic process in the sudden change limit, we find
 \begin{equation}
 \langle W\rangle_{\text{SL}}= \frac{\omega_{f}^2-\omega_i^2}{2\beta \omega_i^2}
  \end{equation}
  and a larger work variance
  \begin{equation}
  \sigma(W)_{\text{SL}}=\frac{1}{\sqrt{2}}\left(\frac{\omega_f}{\omega_i} +1\right) \sigma(W)_{\text{STA}}.
  \end{equation}
Certainly, suppressing the probabilities for
small and large $W$ can be more significant than what is manifested in $\sigma(W)$.
To stress this point,  we display  in Fig.~\ref{JarzynskiPlot} a typical result converging towards Jarzynski's equality with a limited number of classical simulation trajectories.
It is seen that  $\langle e^{-\beta W}\rangle$ in either the bare protocol or the STA converges towards the same theoretical value, but the STA case {does} converge faster (statistically).
This is an intriguing consequence of work fluctuation suppression. { Indeed, the dissipated work, i.e., $\langle W \rangle - \Delta F$
is also suppressed via STA.
STA hence mimic the so-termed escorted free energy simulations put forward in Ref.~\cite{escorted}.
Note, however, that here we use a strictly Hamiltonian control term $ H_{C}(t)$ which not only renders the validity of the Jarzynski
equality but (in contrast to Ref.~\cite{escorted}) leaves the expression for physical work unchanged as well.


\begin{figure}[here]
\resizebox *{6.8cm}{5.1cm}{\includegraphics*{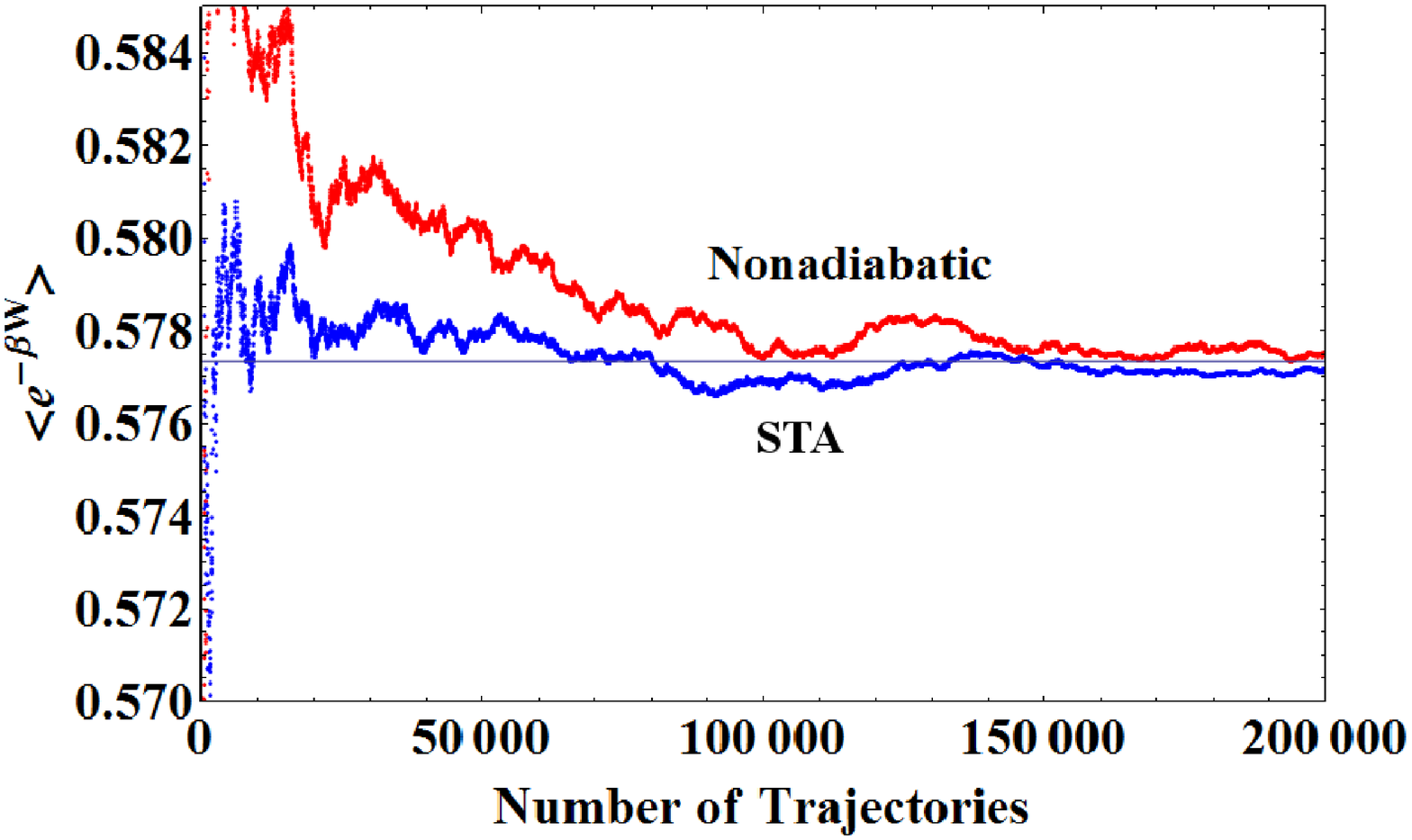}}
\caption{(color online) Numerical average value of $\langle e^{-\beta W}\rangle$ vs the number of classical trajectories used. $\omega_i = 10$, $\beta =0.2$ and $\omega_i \tau= 0.001$.  Upper red (lower blue) line is for a bare nonadiabatic process (STA). Horizontal thin line indicates the theoretical value $1/\sqrt{3}$. The protocol for $\omega(t)$ is the same as in Fig.~1.}
\label{JarzynskiPlot}
\end{figure}



Let us now turn to the quantum work fluctuations with the initial quantum state at thermal equilibrium. Using Eq.~(\ref{quantumP(W)}) one finds that in STA or normal adiabatic processes,
\begin{equation}
P^q_{\rm ad}(W)=\sum\limits_{n=0}^\infty  P_n\delta [ W-\hbar(\omega_{f}-\omega_i) (n+1/2)],
 \end{equation}
with
\begin{equation}
P_n =(1-e^{-\beta\hbar\omega_i }) e^{-n\beta \hbar\omega_i}.
  \end{equation}
  The discrete sum is due to quantization. For general nonadiabatic processes without
$\hat H_C$, analytical but rather complicated work functions are available ~\cite{LutzQ,Lutz2,HanggiNJP}. Here
we {perform} direct numerical investigations with a specific realization of $\omega(t)$ considered in Figs.~\ref{fig:subfig:P(W)} and \ref{JarzynskiPlot}.

\begin{figure}[t]
\begin{center}
\resizebox *{6.8cm}{5.1cm}{
\begin{overpic}[scale=1]{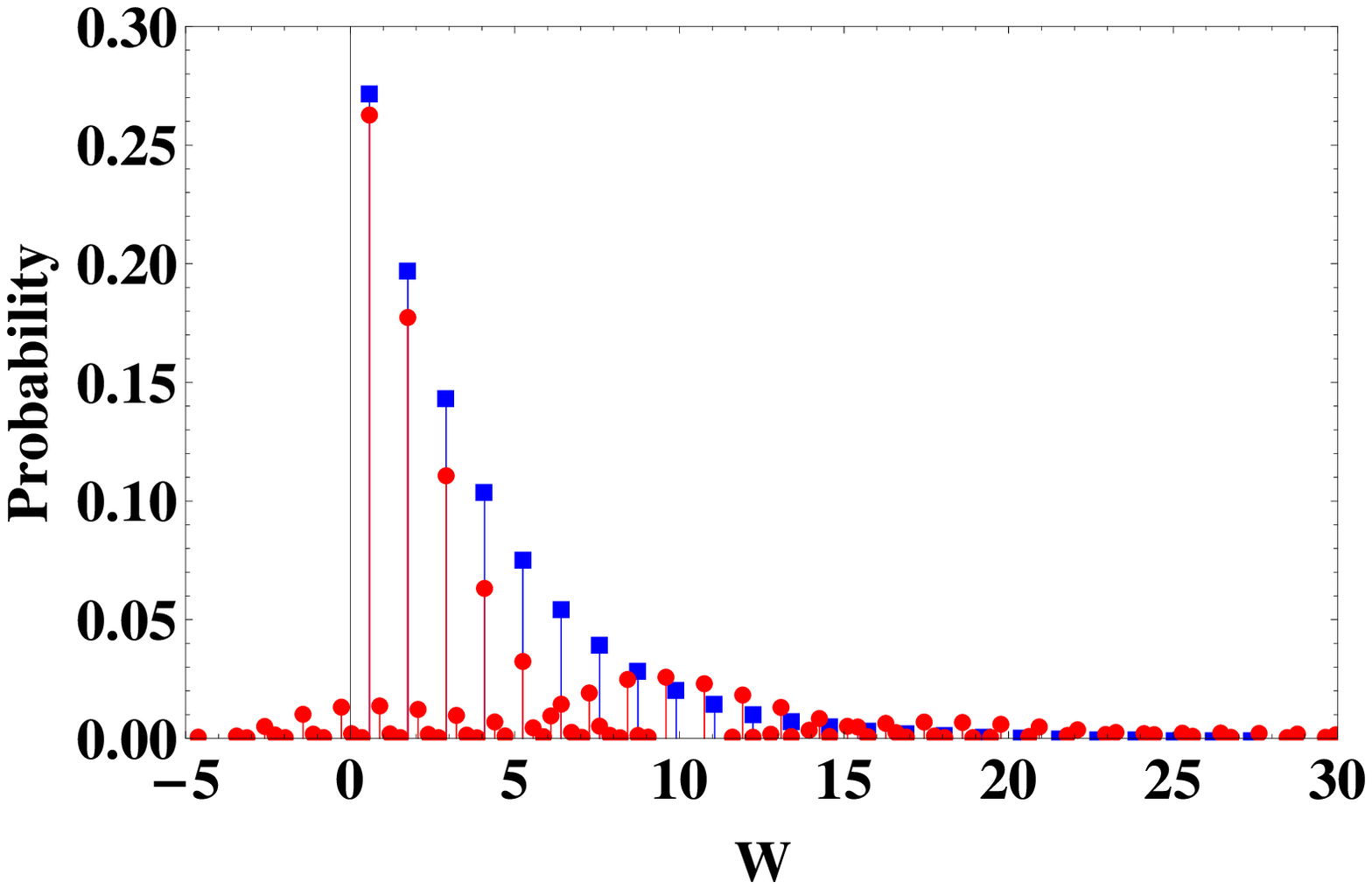}
  \put(38,24){%
    \includegraphics[scale=.8]%
    {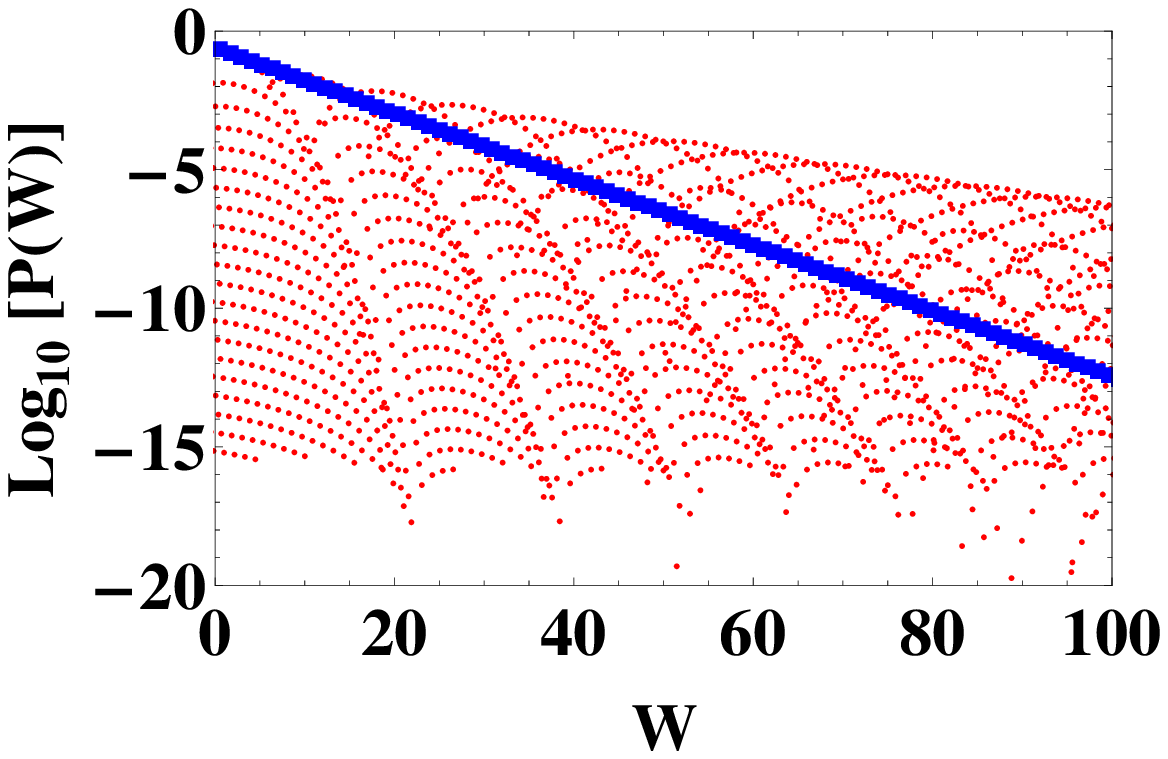}}
\end{overpic}}
\end{center}
\caption{ (color online) Quantum work function from direct simulations, with $2\pi\hbar=1$, $\beta=0.2$. $\omega(t)$ is the same as in Fig.~2. Red dots (blue squares) denote $P^q(W)$ of a bare non-adiabatic process (STA)
with $\tau\omega_i = 0.001$. For clarity only data points
with $P^q(W)\geq 0.0002$ are plotted.  Inset using a semi-log plot displays all data points using a wider range of $W$, with the thick line for STA.}
\label{quantumP(W)plot}
\end{figure}


 {Our results} in Fig.~\ref{quantumP(W)plot} show that the long tail distribution of quantum work is also strongly suppressed in STA, with the degree of suppression in good correspondence with our classical results.
  {Quantitatively, for the parameters in Fig.~3,
the work variance $\sigma(W)=3.1$ ($\sigma(W)=8.7$) in the STA (bare nonadiabatic) case.}
   There are two other {inspiring} aspects.
   First, in the classical protocol (STA or a bare process)
and in the quantum STA, $W$ is {strictly positive-definite with monotonically increasing $\omega(t)$}.
   {In distinct contrast, however,} in a bare quantum processes (i.e. without $\hat H_C$), {work $W$ attains an appreciable probability of becoming negative.}
 Second, {while $W$ for STA assumes  equally spaced, discretized values only,} in the bare processes {work $W$ assumes a rich variety of different discrete values.}
 It is also checked that via quantum Jarzynski's equality the bare protocol and the
STA indeed yield the same quantum $\Delta F$.
%

\section{Enhancement in efficiency and power of a prototype micro heat engine}
We now consider an Otto-cycle based heat engine using a parametric oscillator as its working medium \cite{kosloff,Lutz}. This involves
$\omega$-changing strokes without a reservoir (steps 1, 3) and two relaxation processes with reservoirs
at two different inverse temperatures $\beta_1$ and $\beta_2$ (steps 2, 4).  For the details of the four steps,  see, e.g., Ref.~\cite{Lutz}.

It is convenient to assume steps 1 and 3 to be conventional (quasi) adiabatic or extremely nonadiabatic (sudden change limit) processes.  With $\beta_1$, $\beta_2$, $\omega_i$ and the duration of each cycle step fixed, $\omega_f$ can be optimized to maximize the net work output, with the corresponding engine efficiency being the efficiency at maximum power \cite{curzon}, as denoted $\eta$.  For parameters in the classical regime, $\eta$ obtained using (quasi) adiabatic steps 1 and 3, which is
\begin{equation}
\eta_{\text{ad}}=1-\sqrt{\beta_2/\beta_1},
 \end{equation}
  is higher than (by more than two times)
  \begin{equation}
  \eta_{\text{nonad}}=\frac{1-\sqrt{\beta_2/\beta_1}}{2+\sqrt{\beta_2/\beta_1}}
   \end{equation}
   which is obtained using steps 1 and 3 in their sudden change limit \cite{kosloff,Lutz}. Nevertheless, conventional (quasi) adiabatic processes lead to a very long cycle time, which yields a small engine power. On the other hand, rapid steps 1 and 3 can generate higher power output but
with the drawback of yielding a low engine efficiency $\eta_{\text{nonad}}$.


The balance between power and efficiency is of vast interest to heat engine designs \cite{kosloff,campoPRL}. Here we apply
classical STA to the above-described Otto cycle in the classical domain. That is, we replace steps 1 and 3 by two classical
STA.  In principle, steps 1 and 3 can now be almost instantaneous, but the work function remains to
be identical with that obtained in conventional adiabatic processes. {With the cycle duration now being  bounded only by the relaxation time scales associated with steps 2 and 4, the power of the heat engine is also drastically enhanced}.  We stress that the engine efficiency remains to be $\eta_{\text{ad}}$ simply {because the work functions of steps 1 and 3 are still adiabatic work functions.}
A heat engine based on STA can hence exploit {\it both} advantages of fast strokes (high power) and adiabatic processes (high efficiency).

\begin{figure}[here]
\resizebox *{6.8cm}{5.1cm}{\includegraphics*{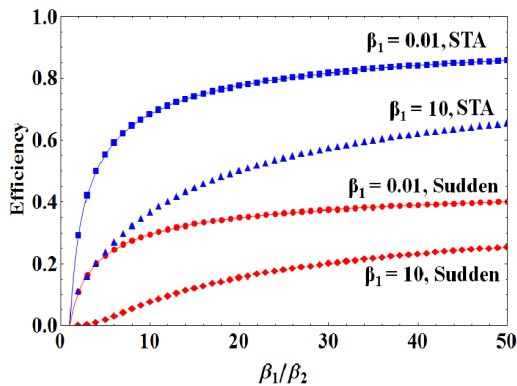}}
\caption{(color online) Efficiency at maximized work output as a function of $\beta_1/\beta_2$ for a prototype quantum heat engine \cite{kosloff,Lutz},
with $\omega_i=10$, $2\pi\hbar=1$, and $\beta_1=10$ or $\beta_1=0.01$.  The two solid lines describe the classical results $\eta_{\text{ad}}$ and $\eta_{\text{nonad}}$ given in the text.  Note the efficiency increase by STA.
}
\label{eng2}
\end{figure}

For the heat engine operating in the quantum regime, we consider quantum STA for steps 1 and 3.
Though the engine power is enhanced,
the expression of quantum work output per cycle remains the same as in Ref.~\cite{Lutz}.  The efficiencies with optimized $\omega_f$ are shown in Fig.~\ref{eng2}.
For a large $\beta_1$ such as $\beta_1=10$,
the obtained efficiency results can be far off from the two classical curves (solid lines) describing $\eta_{\text{ad}}$ and $\eta_{\text{nonad}}$. This case hence represents an engine operating in the deep quantum domain.
Interestingly, there the heat engine efficiency associated with STA is still
more than twice of that obtained in the sudden change limit.  The enhancement factor is even much larger if $\beta_2$ approaches $\beta_1$.


\section {Conclusion}
{Shortcuts to adiabatic processes are shown to have simple classical counterparts. They can substantially suppress work fluctuations in rapid processes.
Work values, though classically positive-definite in a protocol, may be still negative in the quantum domain. However, quantum STA may completely suppress the negative
work values. STA can also enhance the efficiency and at the {\it same} time as well the power of a micro heat engine, in both classical and quantum regimes}.
 Finally, we compare our work with two related and independent studies \cite{campo,chris}. { Reference \cite{campo} studied
 the use of STA {(in a quantum framework)} in a heat engine model, via a tailored time-dependence of the frequency of the parametric oscillator setup \cite{Chen2}.  Though using a different control scheme, the main conclusion drawn in Ref.~\cite{campo} echoes with
 ours in the heat engine application.  Reference \cite{chris} also {studied} the classical analog of quantum STA, but using concepts and techniques different from this work.}

\appendix

\section{$\hat{H}_C$ for Quantum Parameteric Oscillator}
Here we provide some necessary details regarding the quantum
control Hamiltonian for realizing STA.
In our main text we consider a quantum parametric oscillator, whose Hamiltonian is given by
 $\hat H_0(t) = {\hat p^2\over 2m} + {m\over 2} \omega^2(t)\hat q^2$ with a time dependent angular frequency $\omega(t)$.
The calculations for $\hat{H}_C$ can be found from Ref.~\cite{CampoJPB}. For a self-containing comparison with our classical
theory, we first perform a similar calculation here.  Obviously
\begin{equation}
E_n-E_m = \hbar \omega (n-m),
\end{equation}
and
\begin{equation}
\partial_t \hat H_0= m \dot\omega \omega \hat q^2.
\label{appendixA1}
\end{equation}
Note also that
\begin{equation}
\hat q = \sqrt{ \hbar \over 2m\omega}(\hat a+\hat a^{\dag} ),
\end{equation}
where $\hat a$, $\hat a^\dag$ are annihilation and creation operator.  The quantum transitionless driving Hamiltonian derived in Ref.~\cite{Berry} is
\begin{eqnarray}
\hat H_C(t) & = & \img\hbar \frac{\partial U(t,0)}{\partial t} U^{\dagger}(t,0)- \hat H_0[\lambda(t)] \nonumber \\
&=& \img \hbar \sum\limits_{n} \sum\limits_{m\neq n} {|m\rangle \langle m|\partial_t \hat H_0[\lambda(t)] |n\rangle\langle n| \over E_n - E_m},
\label{eqn:quantum_HC}
\end{eqnarray}
from which we have
\begin{eqnarray}
\hat H_C(t) &=& \img \hbar \sum_{i} \sum_{j\neq i} {|j\rangle \langle j|\partial_t \hat H_0(t) |i\rangle\langle i| \over E_i - E_j} \nn
  &=&{\img\hbar \dot\omega \over 2\omega}\sum\limits_{i} \sum\limits_{j\neq i} {|j\rangle \langle j|(\hat a^2+\hat a^{\dag 2} + \hat a \hat a^\dag +\hat a^\dag \hat a) |i\rangle\langle i| \over i - j} \nn
  &=&{\img\hbar \dot\omega \over 2\omega}\sum\limits_{i} {1\over 2 }\Big(\sqrt{(i-1)i}\ |i-2\rangle\langle i|\nn
  &&\qquad -\sqrt{(i+1)(i+2)}\ |i+2\rangle \langle i|\Big) \nn
  &=&{\img\hbar \dot\omega \over 2\omega}\sum\limits_{i} {1\over 2 }(\hat a^2-\hat a^{\dag 2})|i\rangle \langle i| \nn
  &=& -{ \dot\omega \over 4\omega}(\hat q\hat p + \hat p\hat q).
\end{eqnarray}
This is just the expression of $\hat H_C(t)$ given in Ref.~\cite{CampoJPB}.

\section{$H_C$ for Classical Parametric Oscillator}
For a classical parametric oscillator $H_0={ p^2\over 2m} + {m\over 2} \omega^2(t) q^2$, we have \cite{Goldstein}
\begin{equation}
H_0(p,q,\omega)=\tilde H_0(I,\omega) = \omega I
\end{equation}
 In addition, a type-II generating function relates $(p,q)$ and $(I,\theta)$ through
\begin{eqnarray}
p = {\partial F_2(I,q,\omega)\over \partial q}
\label{appendixB1}
\end{eqnarray}
and
\begin{eqnarray}
\theta = {\partial F_2(I,q,\omega)\over \partial I}.
\label{appendixB2}
\end{eqnarray}
Equation (\ref{appendixB1}) leads to
\begin{eqnarray}
\tilde H_0 = \omega I &=&{1\over 2m}\left({\partial F_2(I,q,\omega)\over \partial q}\right)^2+{1\over  2}m\omega^2 q^2, \nn
F_2(I,q,\omega) &=& \int \sqrt{2m\omega I - m^2\omega^2 q^2}dq.
\end{eqnarray}
Therefore
\begin{eqnarray}
\theta = {\partial F_2(I,q,\omega)\over \partial I} &=&\arcsin\left(\sqrt{m\omega\over 2I} q\right)+{\rm const}, \nn
q &=& \left(\sqrt{2I\over m\omega}\right)\sin(\theta-{\rm const}),
\label{appendixB3}
\end{eqnarray}
where ${\rm const}$ is arbitrary and we set it to $0$, which then gives
\begin{equation}
p = \sqrt{2m\omega I - m^2\omega^2 q^2}=\sqrt{2m\omega I}\cos\theta.
\label{appendixB4}
\end{equation}
The control Hamiltonian $K_C$ defined in the main text is then given by
\begin{eqnarray}
K_C[I,\theta,t] &=&- \left[\left( {\partial F_2(I,q,\lambda)\over \partial t}\right)\Biggr|_{I,\ q}\dot \lambda\right]\Biggr|_{q=q(I,\theta,\lambda)} \nn
&=&-{\dot \omega I\over 2\omega}\left[\sin \left(2\arcsin\left(\sqrt{m\omega\over 2I}q\right) \right)\right] \Biggr|_{q=q(I,\theta,\omega)}\nn
&=&-{\dot \omega I\over 2\omega}\sin(2\theta).
\label{appendixB5}
\end{eqnarray}
As such, one directly has $H_C(p,q)=-\frac{\dot\omega}{2\omega}pq$ as given in the main text.

\section{Work Function in Bare Non-Adiabatic Processes}
To compare with the work function in classical STA,  we also present analytical results
for the work functions in classical (bare) non-adiabatic processes, using a parametric oscillator as a specific system.
To that end we first need to specify $\omega(t)$.
In particular, we consider a positive-valued angular frequency
\begin{equation}
\omega(t) =\omega_i \sqrt{{a^2+1\over 2}-{a^2-1\over2}\cos (\pi{t\over\tau})},
\label{appendixC1}
\end{equation}
so that $H_C\propto \dot\omega=0$ at $t=0,\ \tau$.  For brevity we assume $a\equiv \omega_f/\omega_i>1$, i.e. $\omega$ is increasing during the entire protocol.  Then, we have
\begin{eqnarray}
W&=&\int^{\tau}_{0} \frac{dH}{dt} dt = \int ^{\tau}_{0}\frac{\partial H}{\partial t} dt \nonumber\\
&=& \int ^{\tau}_{0} 2\dot{\omega}(t)\omega(t) q^2 dt,
\end{eqnarray}
which clearly indicates that the classical work for a  protocol using a monotonically increasing $\omega(t)$ is positive-definite. Interestingly, as seen in the main text, this is not true in the quantum case.

The equation of motion is
\begin{equation}
\ddot q(t) + \omega^2(t)q(t) =0.
\label{appendixC2}
\end{equation}
There are two linearly independent special solutions $C$ and $S$ with
\begin{eqnarray}
C(0) = 1, \ \ &\ & \ \dot C(0)=0\ ; \nn
S(0) = 0, \ \ &\ &\ \dot S(0)=1.
\label{appendixC3}
\end{eqnarray}
Under our specific choice of $\omega$ in Eq.~(\ref{appendixC1}) (see also Ref.~\cite{Lutz2}), $C$ and $S$ are the basic solutions to Mathieu's equation \cite{NIST}:
\begin{eqnarray}
C(t) &=& w_{\rm I}({\pi\over2\tau}t;c,d); \nn
S(t) &=& {2\tau \over \pi}w_{\rm II}({\pi\over2\tau}t;c,d),
\end{eqnarray} where $c=2\tau^2 \omega_i^2 (a^2+1)/\pi^2$ and $d =\tau^2 \omega_i^2 (a^2-1)/\pi^2 $.
Then a general solution $q(t)$ with initial condition $(p_0,q_0)$ is given by
\begin{eqnarray}
q(t) = q_0 C(t) + {p_0\over m} S(t).
\end{eqnarray}
Once we have the solution $q(t)$, our following derivations will be quite general.

The work done during a time duration $\tau$ is
\begin{eqnarray}
W_\tau &=& H_0[p(p_0,q_0,\tau),q(p_0,q_0,\tau),\tau] - H_0[p_0,q_0,0] \nn
       &=& \cK {\beta\over 2m}p_0^2 + \cL {\beta m\omega_0^2\over 2}q_0^2 +\cM\beta\omega_0 p_0 q_0 \nn
       &=&\cK p'^2 + \cL q'^2 +2 \cM p'q'
       \label{appendixC4}
\end{eqnarray}
where $p' \equiv \sqrt{\beta\over 2m}p_0 $, $ q' \equiv \sqrt{\beta m\omega_0^2\over 2}q_0$, and
\begin{eqnarray}
\cK &\equiv &{1\over \beta} {\left[\dot S^2(\tau) +   \omega_{f}^2 S^2(\tau)-1  \right]}, \nn
\cL &\equiv &{1\over \beta} {\left[{\dot C^2(\tau)\over \omega_0^2} +   {\omega_{f}^2 \over \omega_0^2}C^2(\tau)-1  \right]}, \nn
\cM &\equiv &{1\over \beta \omega_0} {\left[\dot C(\tau) \dot S(\tau)+   \omega_{f}^2 C(\tau)S(\tau)  \right]}.
\end{eqnarray}
Note that $\cK$, $\cL$ and $\cM$ defined above are independent of $(p_0,q_0)$ and hence
 independent of $p'$ and $q'$.
Because $W_\tau$ in Eq.~(\ref{appendixC4}) is expressed in a quadratic form, there always exist a two-dimensional orthogonal transformation such that
\begin{eqnarray}
W_\tau = \mu_+ x^2+\mu_-y^2,
\end{eqnarray}
with
\begin{equation}
\mu_\pm = {1\over 2}\left[(\cK+\cL)\pm\sqrt{(\cK-\cL)^2+4\cM^2}   \right]
\label{appendixC5}
\end{equation}
and $(x,y)$ related to $(p', q')$ by an orthogonal transformation (such that $x^2+y^2=p'^2+q'^2$).
Since the work is known to be positive-definite (under our specific choice of $\omega(t)$), $\mu_\pm$ must be positive.
The initial Gibbs ensemble $\rho_0$ under inverse temperature $\beta$ then becomes
\begin{equation}
\rho_0(p_0,q_0)={1\over Z_0}\nte^{-\beta H_0[p_0,q_0,\omega_0]}={\beta\omega_0\over 2\pi}\nte^{-(x^2+y^2)}.
\end{equation}
Finally, the classical work function is
\begin{eqnarray}
P^c(W) &=& \int\limits_{\Gamma}{\beta\omega_0\over 2\pi}\nte^{-(x^2+y^2)}\delta[W-\mu_+x^2-\mu_-y^2]dp_0dq_0 \nn
       &=& \int\limits_{\Gamma}{1\over\pi}\nte^{-(x^2+y^2)}\delta[W-\mu_+x^2-\mu_-y^2]dx dy \nn
       &=&\int {1\over\pi}e^{-r^2({\cos^2\phi\over\mu_+}+{\sin^2\phi\over\mu_-})}\delta[W-r^2]{r\over\sqrt{\mu_+\mu_-}}drd\phi \nn
     &=&\int_0^{2\pi}{1\over2\pi\sqrt{\mu_+\mu_-}}e^{-W({\cos^2\phi\over\mu_+}+{\sin^2\phi\over\mu_-})}d\phi \nn
     &=&{1\over\sqrt{\mu_+\mu_-}}\exp{\left[-{\mu_++\mu_-\over 2\mu_+\mu_-}W\right]}I_0\left[{\mu_+-\mu_-\over 2\mu_+\mu_-}W\right], \nonumber \\
     \
\end{eqnarray}
where a change of integration variable ($x =r \cos \phi /\sqrt{\mu_+}$ and $y = r \sin \phi /\sqrt{\mu_-})$ with $r\in (0,+\infty)$ and $\phi\in (0,2\pi)$
is used.

\end{document}